\documentclass[12pt]{article}

\usepackage[a4paper,margin=1in]{geometry}
\usepackage{amsmath,amssymb}
\usepackage{hyperref}

\title{Einstein's Worries and Actual Physics: \\Beyond Pilot Waves}

\author{Partha Ghose \footnote{partha.ghose@gmail.com}\\Tagore Centre for Natural Sciences and Philosophy,\\ Rabindra Tirtha, New Town, Kolkata 700156, India}
\date{}

\begin{document}
\maketitle

\begin{abstract}
Tim Maudlin has argued that the standard formulation of quantum mechanics
fails to provide a clear ontology and dynamics and that the de
Broglie--Bohm pilot-wave theory offers a better completion of the
formalism, more in line with Einstein's concerns.  I suggest that while
Bohmian mechanics improves on textbook quantum theory, it does not go
far enough.  In particular, it relies on the ``quantum equilibrium
hypothesis'' and accepts explicit nonlocality as fundamental.  A deeper
completion is available in stochastic mechanics, where the wavefunction
and the Born rule emerge from an underlying diffusion process, and in a
contextual, category-theoretic semantics in which measurement and
EPR--Bell correlations are reinterpreted as features of contextual truth
rather than of mysterious dynamics.  In this framework, the measurement
problem and ``spooky action-at-a-distance'' are dissolved rather than
solved.  Finally, a dynamics based on Rosen's
``classical Schr\"odinger equation'' provides a continuous passage between
quantum and classical regimes, eliminating any sharp Heisenberg cut.
\end{abstract}

\section{Introduction}

Tim Maudlin's recent essay ``Actual Physics, Observation, and Quantum
Theory'' \cite{Maudlin} offers a penetrating analysis of the failure of many physicists
to appreciate the depth of Einstein's criticisms of quantum mechanics.
He is surely right that any acceptable physical theory must provide:
(i) a clear \emph{ontology} (what exists); (ii) a precise \emph{dynamics}
(how it evolves); and (iii) a coherent account of how this ontology and
dynamics connect to \emph{actual observational data}.

Maudlin contends that textbook quantum mechanics, with its dual use of
unitary evolution and an ad hoc collapse postulate, does not meet these
requirements, and that the de Broglie--Bohm pilot-wave theory (Bohmian
mechanics) does \cite{Bohm1952}.  In Bohmian mechanics we have a clear ontology of
particles with definite positions, a deterministic guiding equation for
their trajectories, and an explanation of measurement as nothing over
and above complicated particle dynamics.

I agree with much of this diagnosis. My claim here is that, if one takes
Einstein's demand for a deeper theory seriously, one can and should go
beyond Bohmian mechanics and the very idea of deterministic hidden
variables.  Einstein himself was notably sceptical of such proposals.
In a 1952 letter to Max Born, Einstein commented on Bohm’s deterministic 
hidden-variable theory: ``That way seems too cheap to me.''  

Stochastic mechanics, together with a contextual
semantics for quantum propositions, offers a more radical development of
Einstein's programme, although in a direction he himself might well have
resisted, given its acceptance of ontic randomness and contextuality.
In this approach the wavefunction and the Born rule arise from an
underlying diffusion process---classical in form but intrinsically
stochastic---and the familiar interpretative
problems---measurement, nonlocality, and the role of logic---appear in a
new light.

\section{Bohmian mechanics and quantum equilibrium}

In Maudlin's preferred picture, the basic ontology is simple: point
particles with definite positions in space.  These particles are guided
by a wavefunction $\psi$ that evolves according to the Schr\"odinger
equation.  For a spinless particle of mass $m$, the Bohmian guiding
equation is
\[
  \dot{q}(t) = \frac{\hbar}{m}\,\mathrm{Im}\,\frac{\nabla\psi}{\psi}(q,t),
\]
with an obvious extension to many-particle systems.  Once the initial
positions and the initial wavefunction are fixed, the entire history is
determined.

To recover the empirical success of quantum mechanics, Bohmian mechanics
postulates that the initial distribution of particle positions satisfies
the \emph{quantum equilibrium} condition
\[
  \rho(q,0) = |\psi(q,0)|^2.
\]
One can show that if this holds at one time then it holds at all times,
so that the Born rule is preserved dynamically.  However, quantum
equilibrium is not derived from deeper principles within Bohmian
mechanics; it is an additional assumption about initial conditions, or
justified by further arguments about relaxation and typicality.

Two points follow.  First, the empirical equivalence with standard
quantum mechanics holds only in quantum equilibrium.  Second, there
remains a kind of ``irreducible randomness'' at the level of initial
conditions: we simply assume that Nature picked initial positions
according to $|\psi|^2$.

Moreover, Bohmian mechanics embraces explicit \emph{nonlocality}.  In an
entangled multi-particle state, the velocity of one particle depends
instantaneously on the configuration of others.  Maudlin rightly
emphasizes that Bell's theorem \cite{Bell1964,EPR1935} forces such nonlocality under natural
assumptions; it is not an avoidable blemish.  Still, one may reasonably
ask whether this is the end of the story, especially if one is guided by
Einstein's preference for local fields over action at a
distance.

\section{Stochastic mechanics: from diffusion to Schr\"odinger}

Stochastic mechanics proceeds from a different intuition.  Instead of
postulating hidden variables that evolve deterministically, it posits
that particles undergo an \emph{irreducible Brownian-type motion}, a
physical diffusion superposed on classical forces.  A particle's
position $q(t)$ is described by a stochastic differential equation with
drift and diffusion terms, and its path is continuous but
nowhere-differentiable.

The key result, due to Nelson and others \cite{Nelson1966, Kac1956}, is that under suitable
conditions the probability density $\rho(q,t)$ of this diffusion
process and a phase function $S(q,t)$ can be combined into a complex
field
\[
  \psi(q,t) = \sqrt{\rho(q,t)}\,e^{i S(q,t)/\hbar},
\]
and that $\psi$ satisfies the Schr\"odinger equation, provided the drift
and diffusion are appropriately related.  In the simplest
nonrelativistic case, one finds a relation of the form
\[
  \hbar = m\,\sigma,
\]
where $m$ is the mass of the particles and $\sigma$ is the square root of the diffusion coefficient.  
The wavefunction and the
Born rule then arise together: $\rho$ is the physical probability
density of the diffusion, and $|\psi|^2$ is constructed precisely to
encode it.

From this standpoint, randomness is not merely epistemic ignorance about
hidden variables; it is \emph{ontic}---an intrinsic diffusion in the
dynamics.  At the same time, the randomness is highly structured and
gives rise to the full quantum statistics.  The quantum equilibrium
distribution is no longer a special hypothesis but a manifestation of
the underlying stochastic law.

Bohmian mechanics can be seen, in this perspective, as one way of
repackaging the same statistical content at a more formal level: the
guiding equation is expressed in terms of $\psi$, but the origin of
$\psi$ is left unexplained---the Schr\"{o}dinger equation for $\psi$ is 
accepted as fundamental.  Stochastic mechanics aims to explain it as an emergent entity.

\section{Measurement and the physics of apparatus}
One of Maudlin's most important points is that any adequate theory must
provide a microphysical description of measuring devices.  In special
relativity, we idealize rods and clocks, but we know in principle how to
analyze them as physical systems; there is no deep ``measurement
problem''.  In textbook quantum mechanics, by contrast, the collapse
postulate is an unanalyzed primitive: a rule about what happens when a
``measurement'' occurs, without a corresponding dynamical law for the
apparatus itself.  The theory simply declares that, at some vaguely
defined point in the interaction, the wavefunction of the measured
system is projected onto an eigenstate of the measured observable, with
probabilities given by the Born rule.

From the point of view of ``actual physics'', this is unsatisfactory.
The pointer, the detector, the photoplate are themselves composed of
atoms and fields; in principle they should be describable by the same
dynamical laws as the system.  If one insists on a clean separation
between a quantum system and a classical apparatus, the dynamics across
that separation becomes obscure: the very process that is supposed to
link theory and observation lies outside the theory's own laws.

Stochastic mechanics offers a different route.  A measuring
device---a cloud chamber, a photodetector, a macroscopic pointer---is
treated as a large system of diffusing particles subject to the same
underlying stochastic dynamics as the microscopic system it measures.
The registration of an outcome (a track in a chamber, a macroscopic
current in a wire) is the result of \emph{stochastic amplification} of
microscopic fluctuations, governed by the same diffusion-plus-drift
equations as the rest of the dynamics.  No separate collapse postulate
is needed; there is a single, universal law for the evolution of system
and apparatus together.

In this way, stochastic mechanics realises in concrete form the demand
that the macrophysics of detectors and the microphysics of the systems
they detect should fall under the same theoretical umbrella.  The
measurement problem, in the sense of a fundamental clash between
unitary evolution and collapse, does not arise: the apparent
``collapse'' of probabilities is an emergent feature of stochastic
amplification and contextual coarse-graining, not a basic dynamical
process added by hand.  At the same time, the familiar distinction
between a microscopic ``quantum system'' and a macroscopic measuring
apparatus is recovered as a \emph{dynamical} effect rather than a
principled cut: in the $\sigma$--$\lambda$ scheme (see Section 6) the effective
diffusion coefficient (and hence the strength of the quantum potential)
is significant for small masses and becomes very small for large masses,
so that macroscopic devices behave to high accuracy as classical
pointers, even though they are governed by the same underlying
stochastic dynamics as the microscopic systems they register.

\section{Contextuality, nonlocality, and the role of logic}

So far, the contrast has been framed in terms of dynamics.  But there is
also a logical dimension that Maudlin's essay touches on only indirectly.
Bell's theorem shows that no local hidden-variable theory can reproduce
the quantum correlations if one insists on certain classical assumptions
about joint probabilities.  Bohmian mechanics responds by accepting
nonlocality as a fundamental physical fact.

There is another way to respond: to question the assumed logical
framework.  Quantum phenomena are \emph{contextual}: the truth-values of
propositions about measurement outcomes depend essentially on the
experimental context, and one cannot consistently assign definite
values to all such propositions at once.  This is the content of the
Kochen--Specker theorem and related results.

One can express this mathematically by organizing measurement contexts
into a category and representing context-dependent value assignments as
\emph{presheaves} on that category.  In such a picture, classical
physics corresponds to cases where these presheaves satisfy a sheaf
condition and admit global sections: a single context-independent truth
assignment exists.  Quantum phenomena are precisely those in which there
is no such global section.  Contextuality appears as an obstruction to
global truth, which can be captured, for example, by \v{C}ech
cohomology.

On this reading, the ``spooky action at a distance'' highlighted by Bell
experiments need not be interpreted as a literal superluminal influence.
It is rather the trace of trying to force a \emph{global} Boolean
description onto a world whose structure only supports local,
context-dependent descriptions.  Stochastic mechanics then provides a
natural microphysical substrate for these contextual truth-assignments:
correlations are generated by correlated stochastic dynamics, while the
logical obstruction to global truth reflects the way we organize
experimental contexts.

A fuller development of this sheaf-theoretic reinterpretation of
measurement---including a detailed analysis of presheaves of truth
values, their sheafification, and the role of cohomological obstructions
to global truth---is given in \emph{Measurement as Sheafification:
Context, Logic, and Truth after Quantum Mechanics} \cite{Ghose2025a}.

\section{\texorpdfstring{$\sigma$--$\lambda$}{sigma-lambda} dynamics and the quantum--classical continuum}

The discussion so far has been largely structural.  A natural question
is how the stochastic and contextual picture connects to the familiar
classical limit in a \emph{dynamical} way.  Here a useful starting point
is Rosen's observation that the Schr\"odinger equation, written in polar
form, admits a classical analogue \cite{Rosen1964}.

Writing $\psi = R e^{iS/\hbar}$ and substituting into the Schr\"odinger
equation yields two real equations: a continuity equation for $\rho =
R^2$ and a modified Hamilton--Jacobi equation for $S$,
\[
  \frac{\partial S}{\partial t}
  + \frac{(\nabla S)^2}{2m}
  + V(q,t) + Q[q,t] = 0,
\]
where
\[
  Q[q,t] = -\frac{\hbar^2}{2m}\,\frac{\nabla^2 R}{R}
\]
is the quantum potential.  Rosen proposed a ``classical Schr\"odinger
equation'' obtained, in effect, by setting $Q$ to zero at the
fundamental level and treating $\psi$ as a complex encoding of a
classical ensemble evolving under the ordinary Hamilton--Jacobi dynamics.  In this picture, the quantum potential is what
distinguishes quantum from classical behaviour.

In the stochastic mechanics framework, the quantum potential can be
expressed in terms of the underlying diffusion coefficient $\sigma$,
using the relation $\hbar = m\sigma$.  One can then introduce a
dimensionless parameter $\lambda \in [0,1]$ that scales the strength of
the quantum potential:
\[
  \frac{\partial S}{\partial t}
  + \frac{(\nabla S)^2}{2m}
  + V(q,t) + \lambda\,Q[q,t] = 0.
\]
For $\lambda = 1$ one recovers the full quantum dynamics; for
$\lambda = 0$ one obtains Rosen's classical Hamilton--Jacobi equation
for an ensemble.  Intermediate values $0 < \lambda < 1$ describe
\emph{mesoscopic} regimes in which the quantum potential is partially
effective: the dynamics is neither fully classical nor fully quantum.

If one relates $\lambda$ to the diffusion strength $\sigma$ via
\[
  \lambda = \frac{m\sigma}{\hbar},
\]
or more generally by a monotone function of $m\sigma/\hbar$, then
variations in the underlying stochasticity continuously tune the system
between classical and quantum behaviour.  There is no sharp boundary, no
fundamental ``Heisenberg cut'' separating a microscopic quantum domain
from a macroscopic classical one.  Instead, the emergence of classical,
sheaf-like behaviour (with global sections and approximately Boolean
logic) is tied to the regime in which $\lambda$ is effectively small and
the effect of the quantum potential term becomes negligible compared to the classical terms.

This $\sigma$--$\lambda$ dynamics thus complements the sheaf-theoretic
semantics of measurement.  The presheaf structure encodes the
contextuality of truth values; sheafification describes the logical
passage to global, classical descriptions; and the $\sigma$--$\lambda$
dynamics provides a physical mechanism for moving between regimes in
which contextual, cohomologically nontrivial behaviour is prominent and
regimes in which it is suppressed.  In Bohmian mechanics, by contrast,
there is no analogous continuous parameter built into the dynamics:
quantum behaviour is governed by the full quantum potential at all
scales, and classicality emerges only through coarse-graining and
environmental decoherence.  The conceptual cut between ``quantum'' and
``classical'' is softened but not eliminated.  In the stochastic
$\sigma$--$\lambda$ framework, by contrast, that cut is replaced by a
continuous deformation of the dynamics itself.

\section{Beyond Bohm: toward a deeper completion}

The contrast with Bohmian mechanics can now be summarized.

\begin{itemize}
  \item There is broad \emph{agreement} with Maudlin that standard
        textbook quantum mechanics is conceptually inadequate, and that
        a satisfactory theory must provide an ontology and dynamics that
        apply equally to microscopic systems and macroscopic apparatus.
  \item Bohmian mechanics improves the situation by supplying such an
        ontology and dynamics, but it relies on the quantum equilibrium
        hypothesis and accepts explicit nonlocality as fundamental.
  \item Stochastic mechanics goes further by deriving Schr\"odinger
        dynamics and the Born rule from an underlying diffusion process.
        The wavefunction and quantum equilibrium distribution are
        emergent rather than primitive.
  \item A contextual, category-theoretic semantics for measurement
        outcomes allows us to reinterpret EPR--Bell correlations and the
        measurement problem as consequences of insisting on global
        Boolean logic where only contextual presheaf data are
        available.  In this setting, ``nonlocality'' and collapse no
        longer signal exotic dynamics but the misapplication of a
        classical logical ideal.
  \item The $\sigma$--$\lambda$ dynamics provides a concrete physical
        mechanism for the continuous passage between strongly quantum
        and approximately classical regimes, thereby removing the need
        for any fundamental Heisenberg cut.
\end{itemize}

In this sense, stochastic mechanics, especially in formulations that
explicitly relate the diffusion strength to the quantum potential via a
parameter $\lambda$, comes closer to the kind of deeper completion
Einstein envisaged: a theory in which quantum phenomena arise from an
underlying stochastic dynamics, and in which the conceptual puzzles of
measurement and nonlocality are dissolved by a more appropriate
logical and semantic structure.

\section{Conclusion}

Maudlin's essay performs an important service by restating Einstein's
criticisms in a sharp and historically sensitive way and by insisting
that quantum theory must be brought into line with the standards of
``actual physics''.  The de Broglie--Bohm theory is a serious and
significant step in that direction.  My suggestion is that if one takes
Einstein's demand for a deeper theory fully seriously, one is naturally
led beyond deterministic pilot waves to a genuinely stochastic and
contextual picture.  In such a picture, the wavefunction is emergent,
the Born rule is a manifestation of underlying diffusion, and the
traditional interpretative problems are recognised as artefacts of
forcing classical logic and global truth onto a fundamentally
contextual quantum world.

At the same time, it must be acknowledged that Einstein himself is
unlikely to have welcomed a framework with genuinely \emph{ontic}
randomness.  His famous remark that ``God does not play dice'' expressed
a deep discomfort with fundamental probabilistic laws.  In earlier work
\cite{Ghose2025b} I have analysed his sharp objections to S.~N.~Bose's
probabilistic law of microscopic matter--radiation interactions \cite{Bose1924} and
argued that, once one carefully distinguishes encounter probabilities
from transition rates, those objections can be reconciled with a
stochastic picture that still satisfies Einstein's correspondence
requirement in the classical limit.  From that vantage point, stochastic
mechanics can be seen as realising, in a more modern setting, the kind
of probabilistic microphysics that Bose anticipated and that quantum
optics has since vindicated, while at the same time addressing the
worries about completeness that motivated Einstein's critique of
quantum theory in the first place.

Categorical quantum mechanics, and in particular the work of Coecke,
Paquette and Pavlovi\'c on classical and quantum structuralism
\cite{CoeckePaquettePavlovic2008}, provides a powerful
process-theoretic reconstruction of quantum theory in terms of symmetric
dagger monoidal categories, with classical data picked out by Frobenius
algebra structure.  The categorical machinery is used there to describe
classical--quantum interaction and classical control within a unified
graphical calculus.  The present approach is complementary: it
organises measurement contexts into a category and uses presheaves,
sheafification and cohomology to analyse the logical structure of
contextual truth and its classical, sheaf-like limit.  Whereas
\cite{CoeckePaquettePavlovic2008} models classicality as an internal
algebraic structure in a process category, classical behaviour here
corresponds to the emergence of global sections and Boolean logic from
context-dependent presheaf semantics, with $\sigma$--$\lambda$ dynamics
providing a continuous passage between the two regimes.

\section{Acknowledgements}
While acknowledging use of AI tools for language polishing, I take full responsibility for the scientific content of the paper.

\end{document}